\documentclass
[showkeys,showpacs,superscriptaddress,preprintnumbers,byrevtex]{revtex4}
\usepackage{amsmath,amsfonts,amssymb,amscd,amsxtra,amsthm}
\usepackage{graphicx}
\usepackage{bm}
\usepackage{epstopdf}
\begin{document}
\preprint{CYCU-HEP-10-05}
\title{Chiral restoration at finite temperature with meson loop corrections}
\author{Seung-il Nam}
\email[E-mail: ]{sinam@cycu.edu.tw,sinam@kau.ac.kr}
\affiliation{Department of Physics, Chung-Yuan Christian University, Chung-Li 32023, Taiwan}
\affiliation{Research Institute of Basic Science, Korea Aerospace University, Koyang, 412-791, Korea}
\author{Chung-Wen Kao}
\email[E-mail: ]{cwkao@cycu.edu.tw}
\affiliation{Department of Physics, Chung-Yuan Christian University, Chung-Li 32023, Taiwan}
\date{\today}
\begin{abstract}
We investigate the pattern of chiral symmetry restoration of QCD for $N_{c}=3$ and $N_{f}=2$ at finite temperature ($T$) beyond the chiral limit. To this end, we employ the instanton-vacuum configuration for the flavor SU(2) sector and the Harrington-Shepard caloron for modifying relevant instanton parameters as functions of $T$. The meson loop corrections (MLC), which correspond to $1/N_{c}$ corrections, are also taken into account to reproduce appropriate $m_{q}$ dependences of chiral order parameters. We compute the chiral condensate as a function of $T$ and/or $m_{q}$. We observe that MLC play an important role to have a correct universality-class behavior of chiral-restoration patterns in this framework, depending on $m_{q}$: Second-order phase transition in the chiral limit, $m_{q}=0$ and crossover for $m_{q}\ne0$. Without MLC, all the restoration patterns are crossover, due to simple saddle-point approximations. It turns out that $T^{\chi}_{c}=159$ MeV in the chiral limit and $T^{\chi}_{c}=(177,186,196)$ MeV for $m_{q}=(5,10,15)$ MeV, using the phenomenological choices for the instanton parameters at $T=0$.
\end{abstract}
\maketitle
\section{Introduction}
The spontaneous breaking of chiral symmetry (SB$\chi$S)
has been one of the most important and intriguing subjects for decades because SB$\chi$S reveals complicated structures of the QCD vacuum.
Its restoration at finite temperature ($T$) and/or baryon chemical potential ($\mu$) can be understood microscopically as the QCD vacuum effect is  diminished and the system changes its nature drastically at a certain $\mu$-$T$ point, as $T$ and/or $\mu$ increase.
The restoration of chiral symmetry can be observed by the changes in the chiral order parameters such as the chiral condensate, the chiral susceptibility, dynamically-generated effective quark mass via SB$\chi$S, weak-decay constant and mass of the Nambu-Goldstone (NG) boson, etc.
In principle, lattice QCD (LQCD) simulation is the most promising method to deal with it, although the sign problem has been a huddle at finite $\mu$. To investigate the chiral restoration at finite $T$ and/or $\mu$, there have been also many effective approaches such as the QCD sum rules~\cite{Hatsuda:1991ez,Klingl:1997kf,Kwon:2008vq}, Nambu-Jona-Lasinio (NJL) model~\cite{Buballa:2003qv,Muller:2010am,Schwarz:1999dj}, Dyson-Schinger method~\cite{Blank:2010bz,Hong:1999fh}, Polyakov-loop augmented NJL model (PNJL)~\cite{Fukushima:2003fw,Ratti:2004ra,Ratti:2005jh,Ghosh:2006qh}, gauge-gravity duality model~\cite{Aharony:2006da,Herzog:2006ra,Kobayashi:2006sb}, hidden-local symmetry (HLS) model~\cite{Harada:2003wa,Brown:2009az,Harada:2003jx}, chiral-perturbation theroy ($\chi$PT)~\cite{Kirchbach:1997rk,Meissner:2001gz,GomezNicola:2004gg}, instanton model~\cite{Diakonov:1988my,Carter:1998ji,Nam:2008bq,Nam:2009nn}, functional renormalization-group method~\cite{Braun:2008pi,Braun:2009si}, and so on.

The patterns of chiral symmetry restoration are intricately linked with the quark mass.
At high $T$, for QCD with two-flavor massless quarks, the associated pattern of chiral restoration is belonged to the universal class of O(4) spin model in three dimension, therefore it is second-order transition. However, with the small quark mass, the second-order transition is replaced by a smooth crossover. On the other hand, there is growing evidence show that at low $T$ the chiral restoration in the chiral limit ($m_q=0$) is first- order transition. It suggests that there exists a tricritical point (TCP) where a line of critical point, so called O(4) line turns into first-order transition as $\mu$ increases and/or $T$ decreases. If the quark masses are nonzero, there should be a critical end point (CEP) in the phase diagram of QCD at which the first-order transition line terminates and followed by the crossover when $T$ increases or $\mu$ decreases.
The positions of CEP and TCP of the chiral phase transition have been attracting a lot of interest recently~\cite{Stephanov:1998dy,Hatta:2002sj,Fodor:2004nz,deForcrand:2006pv,de Forcrand:2002ci,Schaefer:2004en,Schaefer:2006ds}. Moreover, the chiral (scalar) susceptibility, which stands for a response to the explicit breakdown of chiral symmetry by nonzero $m_{q}$ and exhibits the pattern of chiral symmetry restoration, is very sensitive to how the QCD vacuum behaves with respect to $m_{q}$. Hence, it is of great importance to study $m_{q}$ dependence for the pattern of chiral restoration in a sophisticated manner.

In this article, we investigate the pattern of chiral symmetry restoration beyond the chiral limit at finite $T$. For this purpose, we employ an effective action (or thermodynamic potential) derived from the instanton-vacuum configuration~\cite{Schafer:1996wv,Diakonov:2002fq}. From the previous work for $(T,\mu)=0$, based on an effective chiral action via the instanton-liquid model~\cite{Nam:2008bq}, it turned out that the meson loop corrections (MLC), which correspond to the $1/N_{c}$ corrections to leading-order contributions based on mean-field approximations, play a critical role to reproduce appropriate behaviors of the chiral order parameters, such as the scalar susceptibility as a function of $m_{q}$. Moreover, if and only if MLC is applied properly, the effective quark mass, which originates from S$\chi$SB and relates to the chiral condensate, showed comparable $m_{q}$ dependences with those obtained by LQCD simulations~\cite{Nam:2008bq,Goeke:2007bj,Kim:2005jc,Bowman:2005vx}. Because we are interested in computing the chiral order parameters in the presence of nonzero $m_{q}$, it is necessary to include MLC in our calculation.

In addition to MLC, we also take into account $T$ modifications on the instanton parameters which are average (anti)instanton size ($\bar{\rho}$) and inter-(anti)instanton distance ($\bar{R}$). In vacuum, their values are estimated phenomenologically as $\bar{\rho}\approx1/3$ fm and $\bar{R}\approx1$ fm, comparable with those from LQCD simulations. Following the previous work~\cite{Nam:2009nn}, we exploit the Harrington-Shepard caloron, corresponding to the temporally-periodic semi-classic solution of Yang-Mill's action in Euclidean space~\cite{Harrington:1976dj}. The fermionic Matsubara formula is also used to evaluate the $T$ dependence of the chiral order parameters. We compute the chiral condensate as a function of $T$ and/or $m_{q}$, and observe that MLC play an important role to agree with the universality-class behavior of the chiral-restoration patterns in this framework: Second-order phase transition in the chiral limit $m_{q}=0$ and crossover for $m_{q}\ne0$. Without MLC, all the restoration patterns are crossover. We also find that $T^{\chi}_{c}=159$ MeV in the chiral limit and $T^{\chi}_{c}=(177,186,196)$ MeV for $m_{q}=(5,10,15)$ MeV by using the phenomenological choices for the instanton parameters: $\bar{R}\approx1$ fm and $\bar{\rho}\approx1/3$ fm. These values of $T^{\chi}_{c}$ are sensitive to the diluteness of instanton ensemble $\sim1/\bar{R}$. Within possible deviation of the instanton parameters, we can obtain values for $T^{\chi}_{c}$ comparable with those from LQCD simulations.

This article is organized as follows: In Section II, we briefly introduce an effective action derived from the instanton-vacuum configuration, and MLC are added to the saddle-point approximation as the $1/N_{c}$ corrections. All the ingredients discussed in Section II are extended to finite $T$ case. For this purpose, we introduce the  Harrington-Shepard caloron and fermionic Matsubara formula. In Section IV, we present and discuss our numerical results for the chiral order parameters as functions of $T$ as well as $m_{q}$. The final Section is devoted for summary and conclusion.

\section{Effective potential with meson-loop corrections}
\begin{figure}[t]
\includegraphics[width=7.5cm]{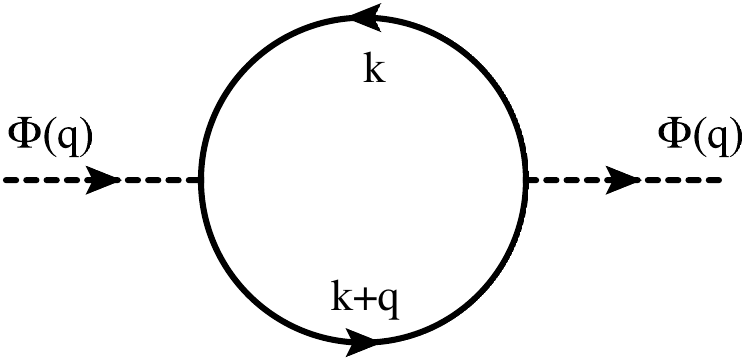}
\caption{An one-loop diagram for the meson correlation function. The solid and dash lines indicate the (anti)quark and meson, respectively.}
\label{FIG0}
\end{figure}

In this Section, we first introduce an effective chiral action derived from the instanton-vacuum configuration, for the leading order of $N_{c}$~\cite{Diakonov:2002fq,Goeke:2007bj,Kim:2005jc,Nam:2008bq}:
\begin{eqnarray}
\label{eq:EA1}
\mathcal{S}_{\mathrm{eff}}&=&\mathcal{C}
+\frac{N}{V}\ln\lambda+2\sigma^{2}
-\int\frac{d^4k}{(2\pi)^4}\mathrm{Tr}
\ln\left[\frac{\rlap{/}{k}+i(m_{q}+M_{k}) }{\rlap{/}{k}+im_{q}}\right],
\end{eqnarray}
where $\mathcal{C}$, $N/V$, and $\lambda$ denote an irrelevant constant for further discussions, instanton number density (instanton packing fraction) being equal to $1/\bar{R}^{4}\approx(1\,\mathrm{fm})^{-4}$, and a Lagrangian multiplier to exponentiate the $2N_{f}$-'t Hooft interacton~\cite{Diakonov:2002fq}, respectively. $\sigma$ indicates the saddle-point value of an isoscalar-scalar meson field, corresponding to SB$\chi$S, and we ignored other meson contributions here. The value of $\sigma$ will be determined self-consistently by solving saddle-point equations in Section IV. Note that this sort of treatment of mesons are equivalent to usual mean-field approximations. $\mathrm{Tr}$ represents a trace over color, flavor, and Lorentz indices. The momentum- and $m_{q}$-dependent effective quark mass originated from nontrivial quark-(anti)instanton interactions is parameterized as
\begin{equation}
\label{eq:MK}
M_{k}=M_{0}\left(\frac{2}{2+\bar{\rho}^{2}k^{2}} \right)^{2}.
\end{equation}
Here, $M_{0}$ and $\bar{\rho}$ stand for effective the quark mass at zero-momentum transfer and average-(anti)instanton size. It is worth mentioning that, from the Fourier transform of the quark zero-mode solution,  the effective quark mass can be written in terms of the modified Bessel functions~\cite{Diakonov:2002fq}. However, we will use the parametrized expression as in Eq.~(\ref{eq:MK}) for numerical convenience. As indicated in Ref.~\cite{Diakonov:2002fq}, as the momentum transfer goes high enough, the parametrization of the effective quark mass becomes different from Eq.~(\ref{eq:MK}) and proportional to $1/k^{6}$. We verified that there appears only negligible difference from using $M_{k}\propto1/k^{6}$ so that we make use of Eq.~(\ref{eq:MK}) for all the momentum-transfer region. Note that in principle $M_{0}$ is a function of $m_{q}$ as well as $T$ implicitly, as will be shown shortly.

If we consider quantum fluctuations around the saddle-point values of the mesons, we can rewrite the effective action with an additional term, which relates to MLC, indicating $1/N_{c}$ corrections of the effective action~\cite{Goeke:2007bj,Kim:2005jc,Nam:2008bq}, as follows:
\begin{eqnarray}
\label{eq:EA3}
\mathcal{S}_{\mathrm{eff}}&=&\mathcal{C}
+\frac{N}{V}\ln\lambda+2\sigma^{2}
-\int\frac{d^4k}{(2\pi)^4}\mathrm{Tr}
\ln\left[\frac{\rlap{/}{k}+i(m_{q}+M_{k}) }{\rlap{/}{k}+im_{q}}\right]
\cr
&+&\frac{1}{2}\sum_{i=1}^{4}\int\frac{d^4k}{(2\pi)^4}
\ln\left\{1-\frac{1}{4\sigma^{2}}
\int\frac{d^4k}{(2\pi)^4}\mathrm{Tr}
\left[ \frac{M_{k}}{\rlap{/}{k}+i(m_{q}+M_{k})}\Gamma_{i}
\frac{M_{k+q}}{\rlap{/}{k}+\rlap{/}{q}+i(m_{q}+M_{k+q})}
\Gamma_{i}\right]\right\},
\end{eqnarray}
where $\Gamma_{i}=(1,\gamma_{5},i\bm{\tau},i\bm{\tau}\gamma_{5})$ relates to the fluctuations from isoscalar-scalar, isoscalar-pseudoscalar, isovector-scalar, and isovector-psuedoscalar mesons. By evaluating over the trace, the effective action can be simplified as
\begin{eqnarray}
\label{eq:EA2}
\mathcal{S}_{\mathrm{eff}}&\approx&\underbrace{\mathcal{C}
+\frac{N}{V}\ln\lambda+2\sigma^{2}-2N_{c}N_{f}
\int\frac{d^4k}{(2\pi)^4}
\ln\left[\frac{k^{2}+\bar{M}^{2}_{k} }
{k^{2}+m^{2}_{q}}\right]}_{\mathrm{LO}}
\cr
&+&\underbrace{\sum_{i=1}^{4}\frac{1}{2}\int\frac{d^4q}{(2\pi)^4}
\ln\left\{1+c^{(1)}_{i}\frac{N_{c}N_{f}}{\sigma^{2}}
\int\frac{d^4k}{(2\pi)^4}
\left[\frac{M_{k}M_{k+q}[k\cdot(k+q)+c^{(2)}_{i}\bar{M}_{k}\bar{M}_{k+q}]}
{(k^{2}+\bar{M}^{2}_{k})[(k+q)^{2}+\bar{M}^{2}_{k+q}]} \right]
\right\}}_{\mathrm{NLO}}
\end{eqnarray}
where we have denoted $1/N_{c}$ leading (LO) and next-leading order (NLO) contributions. Note that we also have used a short-handed notation $\bar{M}_{k}\equiv m_{q}+M_{k}$. As understood in Eq.~(\ref{eq:EA2}), the term inside the square bracket in the last line indicates a (one-loop) correlation function for the relevant mesons, propagating with momentum $q$, as shown in Figure~\ref{FIG0}. The coefficient $c^{(1)}_{i}$ corresponding to each meson is assigned as
\begin{equation}
\label{eq: }
c^{(1)}_{i}=(-1,+1,+3,-3),\,\,\,\,c^{(2)}_{i}=(-1,+1,-1,+1).
\end{equation}

\section{Effective potential at finite temperature}
To investigate the physical quantities at finite $T$, we discuss briefly how to modify the instanton parameters, $\bar{\rho}$ and $\bar{R}$ at finite temperature. We will follow our previous work~\cite{Nam:2009nn} and Refs.~\cite{Harrington:1976dj,Diakonov:1988my} to this end. Usually, there are two different instanton configurations at finite $T$, being periodic in Euclidean time, with trivial and nontrivial holonomies. They are called the Harrington-Shepard~\cite{Harrington:1976dj} and Kraan-Baal-Lee-Lu calorons~\cite{Kraan:1998pm,Lee:1998bb}, respectively. The nontrivial holonomy can be identified as the Polyakov line as an order parameter for the confinement-deconfinement transition of QCD. However, since we are not interested in the confinement-deconfinement phase transition in this work, the Harrington-Shepard caloron is chosen here for simplicity.

\begin{figure}[t]
\begin{tabular}{cc}
\includegraphics[width=8.5cm]{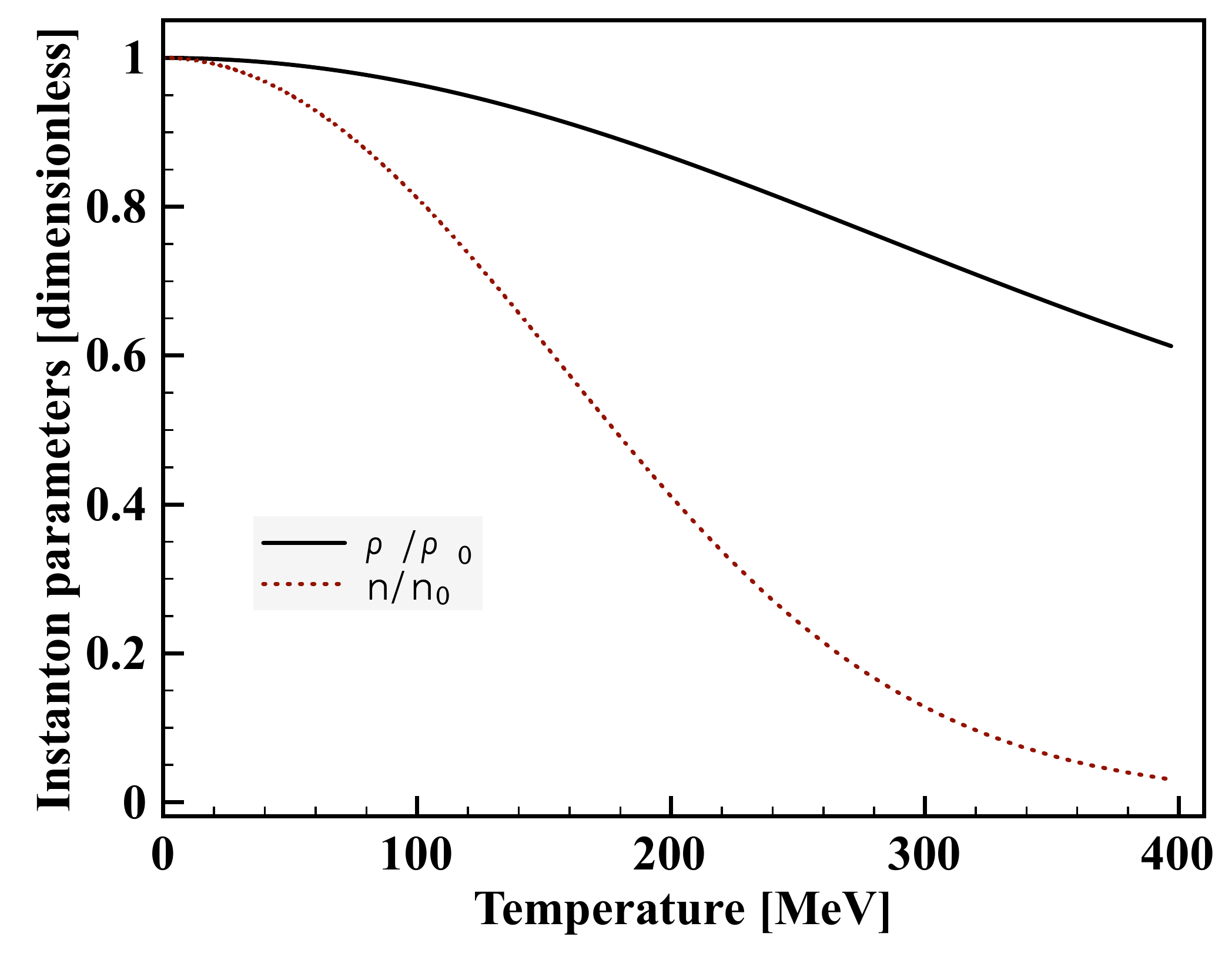}
\includegraphics[width=9.5cm]{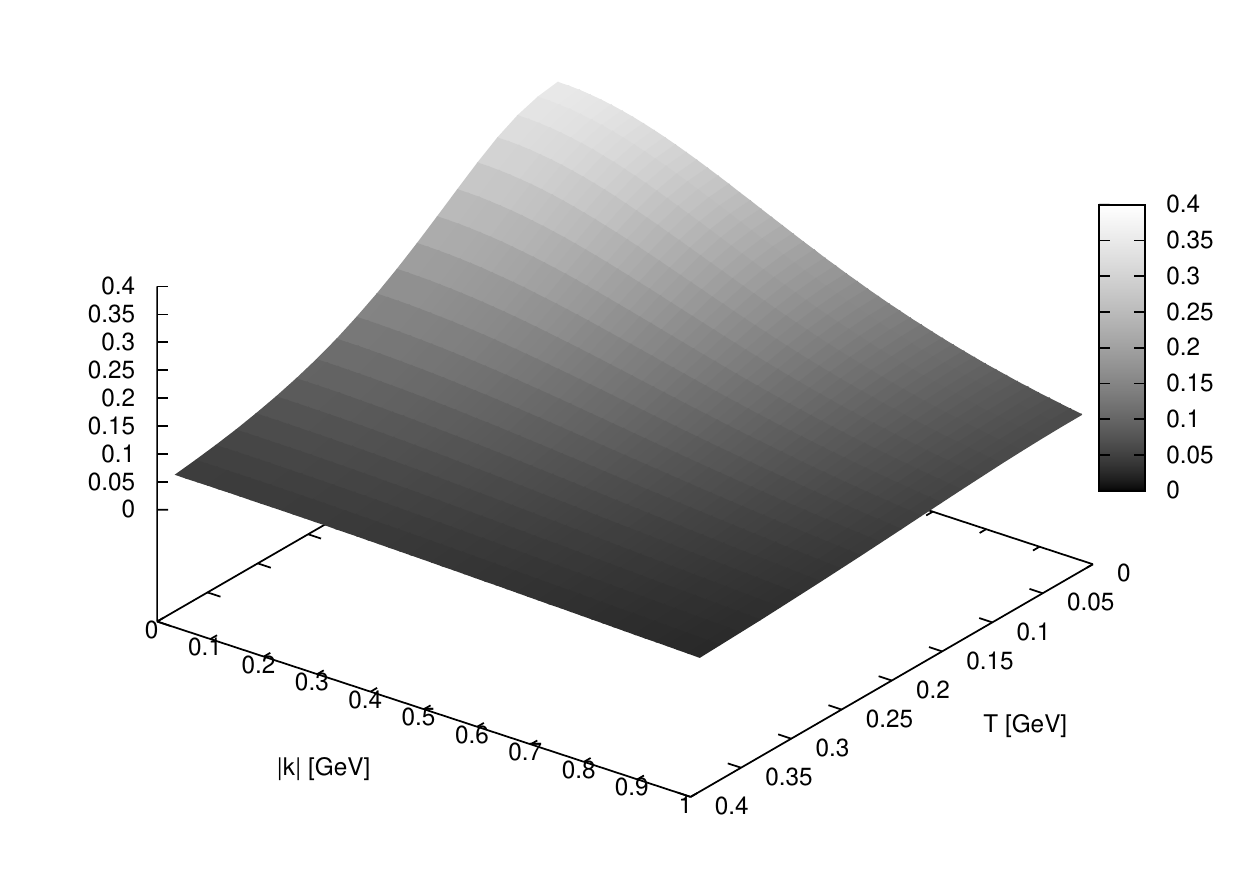}
\end{tabular}
\caption{(Color online) Normalized $\bar{\rho}/\bar{\rho}_{0}$ and $n/n_{0}$ as a function of $T$ for $N_{c}=3$ (left). $M_k$ as a function of $T$ and absolute value of the momentum $|\bm{k}|$ (right).}
\label{FIG1}
\end{figure}

Here, we would like to explain our strategy to modify the effective action in Eq.~(\ref{eq:EA2}) as a function of $T$. As in Ref.~\cite{Nowak:1989jd}, the quark zero-mode solution can be obtained directly by solving the Dirac equation in the presence of the caloron background. By performing a Fourier transform of this zero-mode solution, one is lead to an expression for the $T$-dependent effective quark mass $M_{0}(T)$. However, in the present work, we choose a simpler way to obtain $M_{0}(T)$: Since the effective quark mass can be expressed as a function of the instanton parameters, $\bar{R}$ and $\bar{\rho})$~\cite{Diakonov:2002fq}, instead of solving the Dirac equation directly, we modify $\bar{R}$ and $\bar{\rho}$ as functions of $T$ using the Harrington-Shepard caloron, resulting in the $T$-dependent effective quark mass. As will be shown, this modification together with MLC, we obtain compatible results with that given in Ref.~\cite{Nowak:1989jd} and other similar approaches. We now write the instanton distribution function at finite $T$ with the Harrington-Shepard caloron as follows:
\begin{equation}
\label{eq:IND}
d(\rho,T)=\underbrace{C_{N_c}\,\Lambda^b_{\mathrm{RS}}\,
\hat{\beta}^{N_c}}_\mathcal{C}\,\rho^{b-5}
\exp\left[-(A_{N_c}T^2
+\bar{\beta}\gamma n\bar{\rho}^2)\rho^2 \right].
\end{equation}
Here the abbreviated notations are given as:
\begin{equation}
\label{eq:para}
\hat{\beta}=-b\ln[\Lambda_\mathrm{RS}\rho_\mathrm{cut}],\,\,\,\,
\bar{\beta}=-b\ln[\Lambda_\mathrm{RS}\langle R\rangle],\,\,\,
C_{N_c}=\frac{4.60\,e^{-1.68\alpha_{\mathrm{RS}} Nc}}{\pi^2(N_c-2)!(N_c-1)!},
\end{equation}
\begin{equation}
\label{eq:AA}
A_{N_c}=\frac{1}{3}\left[\frac{11}{6}N_c-1\right]\pi^2,\,\,\,\,
\gamma=\frac{27}{4}\left[\frac{N_c}{N^2_c-1}\right]\pi^2,\,\,\,\,
b=\frac{11N_c-2N_f}{3},\,\,\,\,n=\frac{N}{V}.
\end{equation}
Note that we defined $\hat{\beta}$ and $\bar{\beta}$ at a certain phenomenological cutoff value $\rho_\mathrm{cut}$ and $\langle R\rangle\approx\bar{R}$. Actually only $\bar{\beta}$ is relevant in the following discussions and will be fixed self-consistently within the present framework. $\Lambda_{\mathrm{RS}}$ stands for a scale depending on a renormalization scheme, whereas $V_3$ stands for the three-dimensional volume. Using the instanton distribution function in Eq.~(\ref{eq:IND}), we can compute the average value of the instanton size, $\bar{\rho}^2$ straightforwardly as follows~\cite{Schafer:1996wv}:
\begin{equation}
\label{eq:rho}
\bar{\rho}^2(T)
=\frac{\int d\rho\,\rho^2 d(\rho,T)}{\int d\rho\,d(\rho,T)}
=\frac{\left[A^2_{N_c}T^4
+4\nu\bar{\beta}\gamma n \right]^{\frac{1}{2}}
-A_{N_c}T^2}{2\bar{\beta}\gamma n},
\end{equation}
where $\nu=(b-4)/2$. Substituting Eq.~(\ref{eq:rho}) into Eq.~(\ref{eq:IND}), the distribution function can be evaluated further as:
\begin{equation}
\label{eq:dT}
d(\rho,T)=\mathcal{C}\,\rho^{b-5}
\exp\left[-\mathcal{M}(T)\rho^2 \right],\,\,\,\,
\mathcal{M}(T)=\frac{1}{2}A_{N_c}T^2+\left[\frac{1}{4}A^2_{N_c}T^4
+\nu\bar{\beta}\gamma n \right]^{\frac{1}{2}}.
\end{equation}
The instanton-number density $n$ can be computed self-consistently as a function of $T$, using the following equation:
\begin{equation}
\label{eq:NOVV}
n^\frac{1}{\nu}\mathcal{M}(T)=\left[\mathcal{C}\,\Gamma(\nu) \right]^\frac{1}{\nu},
\end{equation}
where we have replaced $NT/V_3\to n$, and $\Gamma(\nu)$ indicates a $\Gamma$ function with an argument $\nu$. Note that $\mathcal{C}$ and $\bar{\beta}$ can be determined easily using Eqs.~(\ref{eq:rho}) and (\ref{eq:NOVV}), incorporating the vacuum values of the $n$ and $\bar{\rho}$: $\mathcal{C}\approx9.81\times10^{-4}$ and $\bar{\beta}\approx9.19$. Using these results we can obtain the average instanton size $\bar{\rho}$ as a function of $T$ with Eq.~(\ref{eq:rho}).
The $T$ dependences of the normalized $\bar{\rho}/\bar{\rho}_{0}$ and $n/n_{0}$ are plotted in the left panel of Fig.~\ref{FIG1}. As shown there, these quantities are decreasing with respect to $T$ as expected. However, even beyond $T^{\chi}_{c}\approx\Lambda_{\mathrm{QCD}}\approx200$ MeV, the instanton contribution remains finite.

\section{Effective thermodynamic potential with meson-loop corrections}
To use the effective action in Eq.(4) to study the pattern of chiral symmetry restoration, one needs extend it to a finite $T$ case.
For this purpose, we make use of the fermionic Matsubara formula as follows:
\begin{equation}
\label{eq:MA}
\int\frac{d^4k}{(2\pi)^4}\to
T\sum_{n=-\infty}^{\infty}\int\frac{d^3\bm{k}}{(2\pi)^3}f(w_{n},\bm{k})
=T\sum_{n}\int \frac{d^3\bm{k}}{(2\pi)^3}f(w_{n},\bm{k}),
\end{equation}
where $w_{n}$ and $\bm{k}$ stand for the fermionic Matsubara frequency assigned as $(2n+1)\pi T$ for $n\in\mathcal{I}$ and the three momentum of the quark. Substituting Eq.~(\ref{eq:MA}) into Eq.~(\ref{eq:EA2}), one can derive an effective thermodynamic potential for a unit volume as
\begin{eqnarray}
\label{eq:TECA1}
\Omega_{\mathrm{eff}}&\approx&\underbrace{\mathcal{C}
+\frac{N}{V}\ln\lambda+2\sigma^{2}
-2N_{c}N_{f}T\sum_{n}\int \frac{d^3\bm{k}}{(2\pi)^3}
\ln\left[\frac{w^{2}_{n}+\bm{k}^{2}+\bar{M}^{2}_{a}}
{w^{2}_{n}+\bm{k}^{2}+m^{2}_{q}}\right]}_{\mathrm{LO}}
\cr
&+&\underbrace{\frac{\Lambda}{4\pi}\int\frac{d^3\bm{q}}{(2\pi)^3}
\ln\left\{1-\frac{3N_{c}N_{f}T}{\sigma^{2}}
\sum_{n}\int \frac{d^3\bm{k}}{(2\pi)^3}
\left[\frac{M_{a}M_{b}[w_{n}^{2}+\bm{k}\cdot(\bm{k}+\bm{q})
+\bar{M}_{a}\bar{M}_{b}]}
{(w^{2}_{n}+E^{2}_{a})(w_{n}^{2}+E^{2}_{b})} \right]\right\}}_{\mathrm{NLO}}.
\end{eqnarray}
In deriving above effective thermodynamic potential from Eq.~(\ref{eq:EA2}), we have taken the following approximations:
1) For the integral over $q_{4}$, we set a cutoff mass $\Lambda$. 2) We also ignore $q_{4}$ dependence inside the square bracket of the NLO contribution for simplicity. In this way, we have an additional parameter $\Lambda$ in the present framework. However, as shown in Ref.~\cite{Nam:2008bq}, since the isovector-pseudoscalar meson, i.e. pion dominates the meson fluctuations, it is reasonable to set the cutoff $\Lambda$ proportional to $m_{\pi}$ as follows:
\begin{equation}
\label{eq:LAMBDA}
\Lambda\approx m_{\pi}\frac{\bar{\rho}_{0}}{\bar{\rho}}.
\end{equation}
Note that, in the above equation, we have multiplied a factor $\bar{\rho}_{0}/\bar{\rho}$ to $m_{\pi}$ in order to include $T$ dependence of the cutoff mass. Moreover, this multiplication factor represents a correct chiral-restoration pattern of $m_{\pi}$, i.e. the mass of pion, as a NG boson, increases as SB$\chi$S restored. In the left panel of Figure~\ref{FIG1}, this factor increases as $T$ does show a reasonable behavior under chiral restoration transition, i.e. $m_{\pi}$ gets heavier as chiral symmetry is restored. We also have taken into account the third approximation: 3) As mentioned above, the fluctuation from pion dominates the NLO contribution. Hence, we ignored all the meson fluctuations except for that from the pion. According to these approximations 1) $\sim 3)$, we obtain the expression for the thermodynamic potential with the MLC contribution in Eq.~(\ref{eq:TECA1}). The energies for the quarks are written as $E^{2}_{a}=\bm{k}^{2}+\bar{M}^{2}_{a}$ and $E^{2}_{b}=(\bm{k}+\bm{q})^{2}+\bar{M}^{2}_{b}$. Accordingly, the momentum- and $T$-dependent effective quark mass read:
\begin{equation}
\label{eq:Mab}
M_{a}=M_{0}\left[\frac{2}{2+\bm{k}^{2}\bar{\rho}^{2}} \right],
\,\,\,\,
M_{b}=M_{0}\left[\frac{2}{2+(\bm{k}+\bm{q})^{2}\bar{\rho}^{2}} \right].
\end{equation}
Here, we have ignored the fourth component of the momentum for brevity, $k_{4}\to0$, and verified that this simplification does not make considerable difference in comparison to the full calculations. Note that $\bar{\rho}$ inside $M_{a,b}$ in Eq.~(\ref{eq:Mab}).

The value of $M_{0}$ is obtained as a function of $m_{q}$ and $T$ by solving the self-consistent saddle-point equation with respect to $\lambda$:
$\partial\Omega_{\mathrm{eff}}/\partial\lambda=0$, resulting in
\begin{eqnarray}
\label{eq:NOV}
\frac{N}{V}&\approx&
2N_{c}N_{f}T\sum_{n}\int \frac{d^3\bm{k}}{(2\pi)^3}\frac{M_{a}\bar{M}_{a}}
{w^{2}_{n}+E^{2}_{a}}
+\frac{\Lambda}{4\pi}\int\frac{d^3\bm{q}}{(2\pi)^3}
\frac{T\sum_{n}\int \frac{d^3\bm{k}}{(2\pi)^3}\left[
\frac{3M_{a}M_{b}(w^{2}_{n}+\xi^{2})}
{(w^{2}_{n}+E^{2}_{a})(w_{n}^{2}+E^{2}_{b})}\right]}
{T\sum_{n}\int \frac{d^3\bm{k}}{(2\pi)^3}\left[\frac{M_{a}\bar{M}_{a}}
{w^{2}_{n}+E^{2}_{a}}\right]}
\cr
&=&2N_{c}N_{f}\int\frac{d^3\bm{k}}{(2\pi)^3}F_{0}+
\frac{\Lambda}{4\pi}\int\frac{d^3\bm{q}}{(2\pi)^3}
\left[\frac{\int\frac{d^3\bm{k}}{(2\pi)^3}
\left(F_{1}+F_{2} \right)}
{\int\frac{d^3\bm{k}}{(2\pi)^3}F_{0}} \right],
\end{eqnarray}
where we have introduced a notation $\xi^{2}=\bm{k}\cdot(\bm{k}+\bm{q})+\bar{M}_{a}\bar{M}_{b}+(M_{a}\bar{M}_{b}+\bar{M}_{a}M_{b})/2$. The relevant functions $F_{0\sim2}$ after summing over $n$ are given in Appendix. Note that $N/V$ in Eq.~(\ref{eq:NOV}) is a function of $T$ as discussed in the previous section.

We plot the momentum- and $T$-dependent effective quark mass $M_{a}$ in Eq.~(\ref{eq:Mab}) for $m_{q}=0$ in the right panel of Figure~\ref{FIG1}.
As shown in the figure, $M_{a}$ is a decreasing function of $|\bm{k}|$ as well as $T$, according to decreasing instanton effect. Thus, the effective quark mass plays a role of a natural UV regulator and signals the chiral restoration with respect to $T$. Similarly, the $m_{q}$ dependence of $M_{0}$ can be easily computed by solving Eq.~(\ref{eq:NOV}) by putting nonzero $m_{q}$ in a self-consistent manner. We will see the $m_{q}$ dependence of the effective quark mass in detail in the next Section. However, before going further, we want mention that here is one assumption in obtaining $M_{0}$ as a function of $m_{q}$: $\bar{\rho}$ and $\bar{R}$ are not dependent on $m_{q}$ but only $T$. We argue that this assumption is reasonable considering that, since these instanton parameters, indicating the QCD vacuum properties approximately, do not correspond to external sources such as $m_{q}$, they behave independently with respect to $m_{q}$. The chiral condensate, $\langle iq^{\dagger}q\rangle $ is obtained from the thermodynamic potential as follows:
\begin{eqnarray}
\label{eq:CC}
\langle iq^{\dagger}q \rangle&=&
-\frac{1}{N_{f}}\frac{\partial\Omega_{\mathrm{eff}}}{\partial m_{q}}
\approx4N_{c}T\sum_{n}\int \frac{d^3\bm{k}}{(2\pi)^3}
\left[\frac{\bar{M}_{a}}{w^{2}_{n}+E^{2}_{a}}
-\frac{m_{q}}{w^{2}_{n}+E^{2}_{0}}  \right]
+\frac{\Lambda}{4\pi N_{f}}\int\frac{d^3\bm{q}}{(2\pi)^3}
\frac{T\sum_{n}\int \frac{d^3\bm{k}}{(2\pi)^3}\left[
\frac{3M_{a}M_{b}(\bar{M}_{a}+\bar{M}_{b})}
{(w^{2}_{n}+E^{2}_{a})(w_{n}^{2}+E^{2}_{b})}\right]}
{T\sum_{n}\int \frac{d^3\bm{k}}{(2\pi)^3}\left[\frac{M_{a}\bar{M}_{a}}
{w^{2}_{n}+E^{2}_{a}}\right]}
\cr
&=&4N_{c}\int\frac{d^3\bm{k}}{(2\pi)^3}
\left(G_{0}-G_{1}\right)
+\frac{\Lambda}{4\pi N_{f}}\int\frac{d^3\bm{q}}{(2\pi)^3}
\left[\frac{\int\frac{d^3\bm{k}}{(2\pi)^3}G_{2}}
{\int\frac{d^3\bm{k}}{(2\pi)^3}F_{0}} \right].
\end{eqnarray}
Here, $E^{2}_{0}=k^{2}+m^{2}_{q}$ and $G_{0\sim2}$ are also given in Appendix. For more details of derivation of Eqs.~(\ref{eq:NOV}) and (\ref{eq:CC}), one can refer to Ref.~\cite{Nam:2008bq}.
Since the chiral condensate is one of the most pronounced order parameters for the breakdown and restoration of SB$\chi$S, we briefly discuss how we study the chiral-restoration pattern using the chiral condensate.
At certain $T$, SB$\chi$S is restored from the NG phase, following the universality-class pattern of QCD phase transition.  Thus, at the critical $T$ for the restoration, denoted by $T^{\chi}_{c}$, the chiral condensate behaves irregularly. Hence, by seeing this irregularity, we can have information on the chiral-restoriation patterns from the order parameters. Another way is to observe the behavior of the chiral susceptibility with respect to $T$~\cite{Fukushima:2003fw}. In the present work, following Refs.~\cite{Ratti:2004ra,Ratti:2005jh,Rossner:2007ik}, we sort out $T^{\chi}_{c}$ by differentiating the chiral condensate with respect to $T$, $\partial\langle iq^{\dagger}q\rangle/\partial T$. At $T^{\chi}_{c}$, $\partial\langle iq^{\dagger}q\rangle/\partial T$ becomes finite maximum for the crossover and infinity for the (first) second-order phase transitions.
\section{Numerical results}
\begin{figure}[t]
\begin{tabular}{cc}
\includegraphics[width=8.5cm]{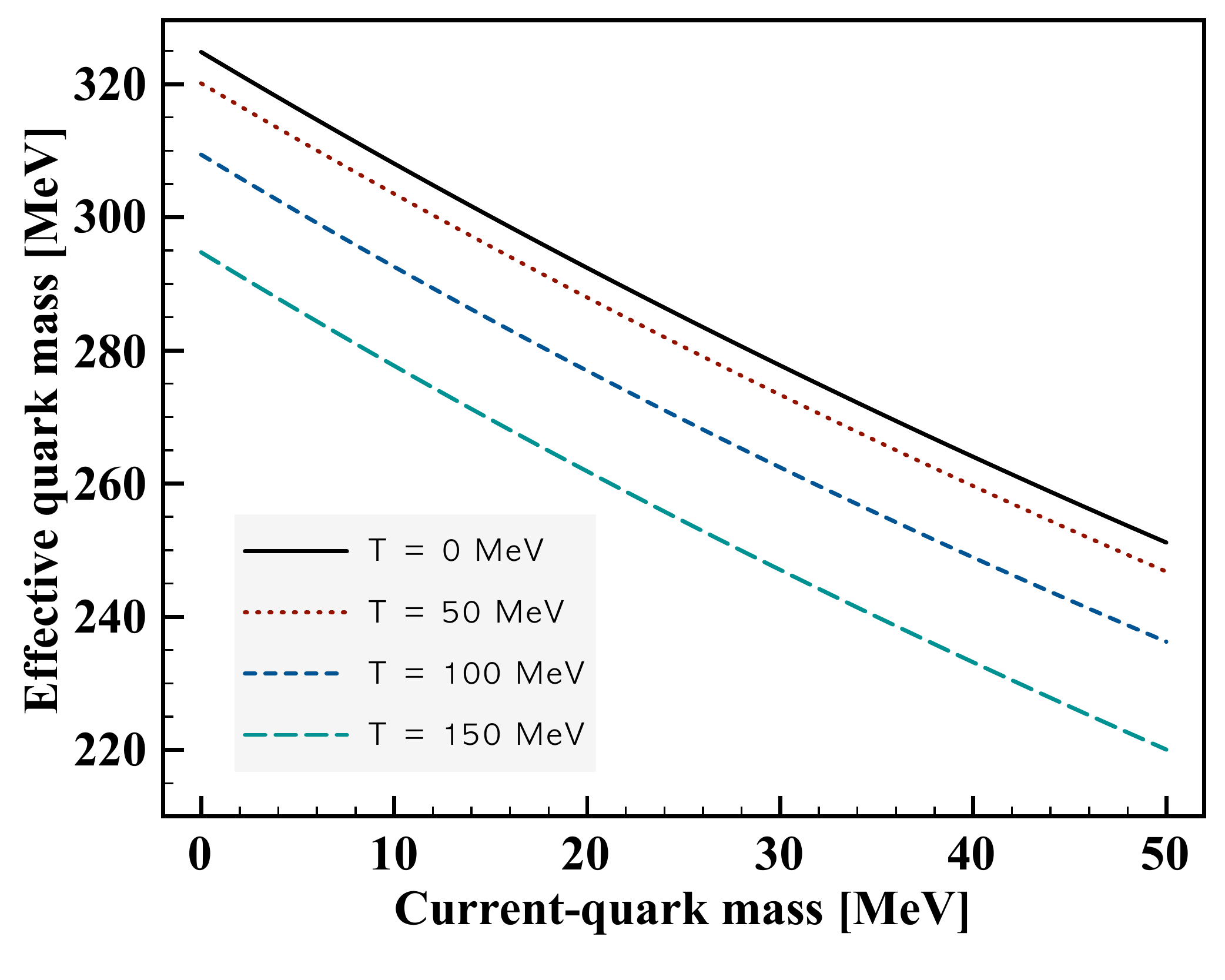}
\includegraphics[width=8.5cm]{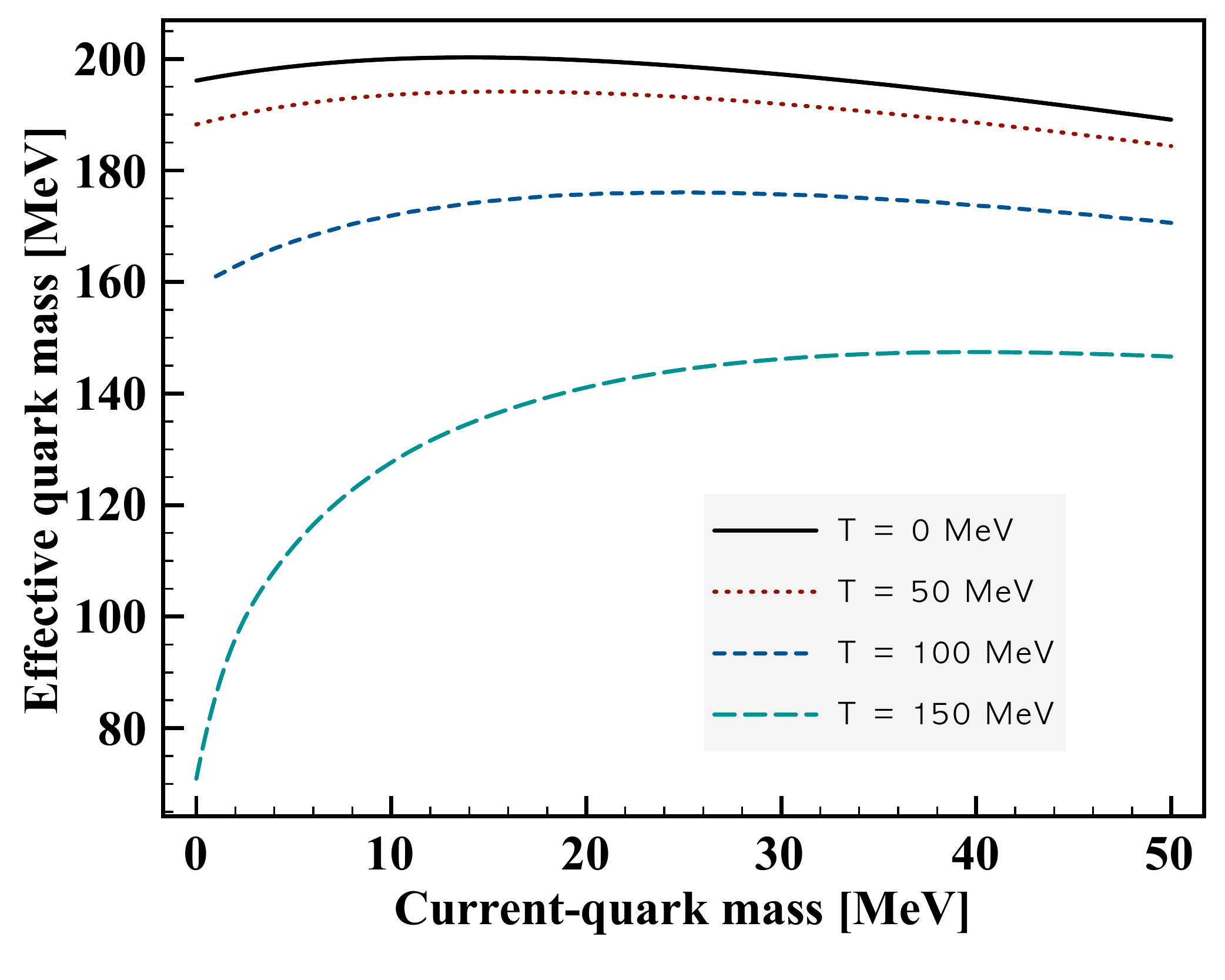}
\end{tabular}
\caption{(Color online) Effective quark mass $M_{0}$ in Eq.~(\ref{eq:Mab}) as a function of current-quark mass $m_{q}$ for the different values of temperature, $T=(0\sim150)$ MeV without (left) and with (right) the meson-loop corrections.}
\label{FIG34}
\end{figure}

In this section, we present the numerical results for various quantities, such as the effective quark mass and the chiral condensate as functions of $m_{q}$ and $T$. To identify the effects from the MLC contribution, we will show those quantities with and without MLC, separately. In Figure~\ref{FIG34}, we depict the effective quark mass with zero-momentum transfer $M_{0}$, which was determined by solving the saddle-point equation in Eq.~(\ref{eq:NOV}), as a function of $m_{q}$ for $T=(0\sim150)$ MeV. Note that the left panel shows it without MLC, while the right one with MLC. As shown in the left panel. $M_{0}$ turns out to be a monotonically decreasing function of $m_{q}$ in the absence of MLC. The value of $M_{0}$ without MLC in the chiral limit is about $325$ MeV and becomes smaller as $T$ increases. This observation is consistent with that given in Ref.~\cite{Musakhanov:2001pc} for the case at $T=0$.

However, the situation changes drastically when MLC is included. At $T=0$, the curve of $M_{0}$ increases
until $m_{q}\approx10$ MeV then starts to decrease after it. In the chiral limit, the value of $M_{0}$ is about $190$ MeV.
It is about $45\%$ less than the value without MLC.
We note that this value of $M_{0}$ is considerably different from that without MLC. Actually, this result agrees with
with the result of the previous work
Refs.~\cite{Goeke:2007bj,Kim:2005jc}. The value of $M_{0}$ is changed from $350$ MeV to $125$ MeV by including
MLC in ~\cite{Goeke:2007bj} and from $567$ MeV to $360$ MeV in ~\cite{Kim:2005jc}. The MLC reduces $M_{0}$ at $T=0$ ranged from $37\%$ to $63\%$.
The difference between the results in \cite{Goeke:2007bj} and \cite{Kim:2005jc} is due to the different instanton parameters $\bar{R}$ and $\bar{\rho}$.  In Refs.~\cite{Goeke:2007bj,Kim:2005jc}, it was also argued that the MLC contribution corresponds to the chiral log term which provides considerable contributions to the chiral condensate~\cite{Novikov:1981xi}. Hence, considering that $\langle \bar{q}q\rangle \propto M_{0}$ in general SB$\chi$S pictures, this drastic changes of $M_{0}$ can be understood in the same way since our present framework is essentially equivalent to that of Refs.~\cite{Goeke:2007bj,Kim:2005jc}. We indeed verify that the value of $M_{0}$ with MLC is very sensitive to the instanton parameters. Furthermore we also verified that the fine tuning of the instanton parameters provides only negligible changes in the value of $T^{\chi}_{c}$. The shape of the curve for $T=0$ is well consistent with the LQCD simulation~\cite{Bowman:2005vx}, although the magnitude of the curve is rather different. The difference between our result and the LQCD simulation can be understood by that the different renormalization scales between them: $0.6$ GeV for ours and $1.65$ GeV for Ref.~\cite{Bowman:2005vx}. Thus, the value of $M_{0}$ is very sensitive to the choice of $\bar{R}$ so that one may obtain more comparable results with the LQCD data by tuning those instanton parameters.

As $T$ goes higher, the curve of $M_{0}$ changes its shape as well as magnitude as shown in the right panel of Figure~\ref{FIG34}. As expected from the decreasing instanton effects with respect to $T$, the magnitude of  the curves becomes smaller being similar to that without MLC. On the contrary, the value of $M_{0}$ in the vicinity of $m_{q}=0$ is more sensitive to $T$ in comparison to that without MLC, whereas the curves become relatively flat as $m_{q}$ increases for $m_{q}\gtrsim10$ MeV for all $T$ values. In other words, focusing on the region $m_{q}\lesssim10$ MeV,  this behavior indicates that the effect of the explicit chiral-symmetry breaking becomes obvious. This manifestation of the explicit chiral-symmetry breaking at small $m_{q}$ can be explained again in terms of the decreasing instanton effects, i.e. decreasing nonperturbative QCD vacuum effects. Since the strength of the instanton effects prevails over that of the explicit chiral-symmetry breaking for lower $T$, one can see only small difference in $M_{0}$ with respect to $m_{q}$, and vice versa for higher $T$. For the larger $m_{q}$ beyond $m_{q}\approx20$ MeV, this manifestation become obscure since $M_{0}$ is almost saturated as already mentioned. This behavior can be also analytically understood by seeing the functions $F_{1}$ and $F_{2}$, which contains ($E^{2}_{a}-E^{2}_{b}$) in the denominator as shown in Appendix. In a simple analysis, this term gives $\mathcal{O}(m^{-1}_{q})$:
\begin{equation}
\label{eq:OOO}
\frac{\mathrm{Func}(T)}{E^{2}_{a}-E^{2}_{b}}\approx
\frac{\mathrm{Func}(T)}{(\Delta k^{2})-2(\Delta M) (m_{q}+M_{0})},
\end{equation}
where $\Delta k^{2}\equiv k^{2}_{a}-k^{2}_{b}$ and the same for $\Delta M$, and $\mathrm{Func}(T)$ is an appropriate function of $T$. For $\Delta k^{2}>0$, we have $\Delta M>0$. Hence, the functions $F_{1,2}$ are enhanced for smaller $m_{q}$, enhancing the $T$ dependence, $\mathrm{Func}(T)$, simultaneously.

\begin{figure}[t]
\begin{tabular}{cc}
\includegraphics[width=8.5cm]{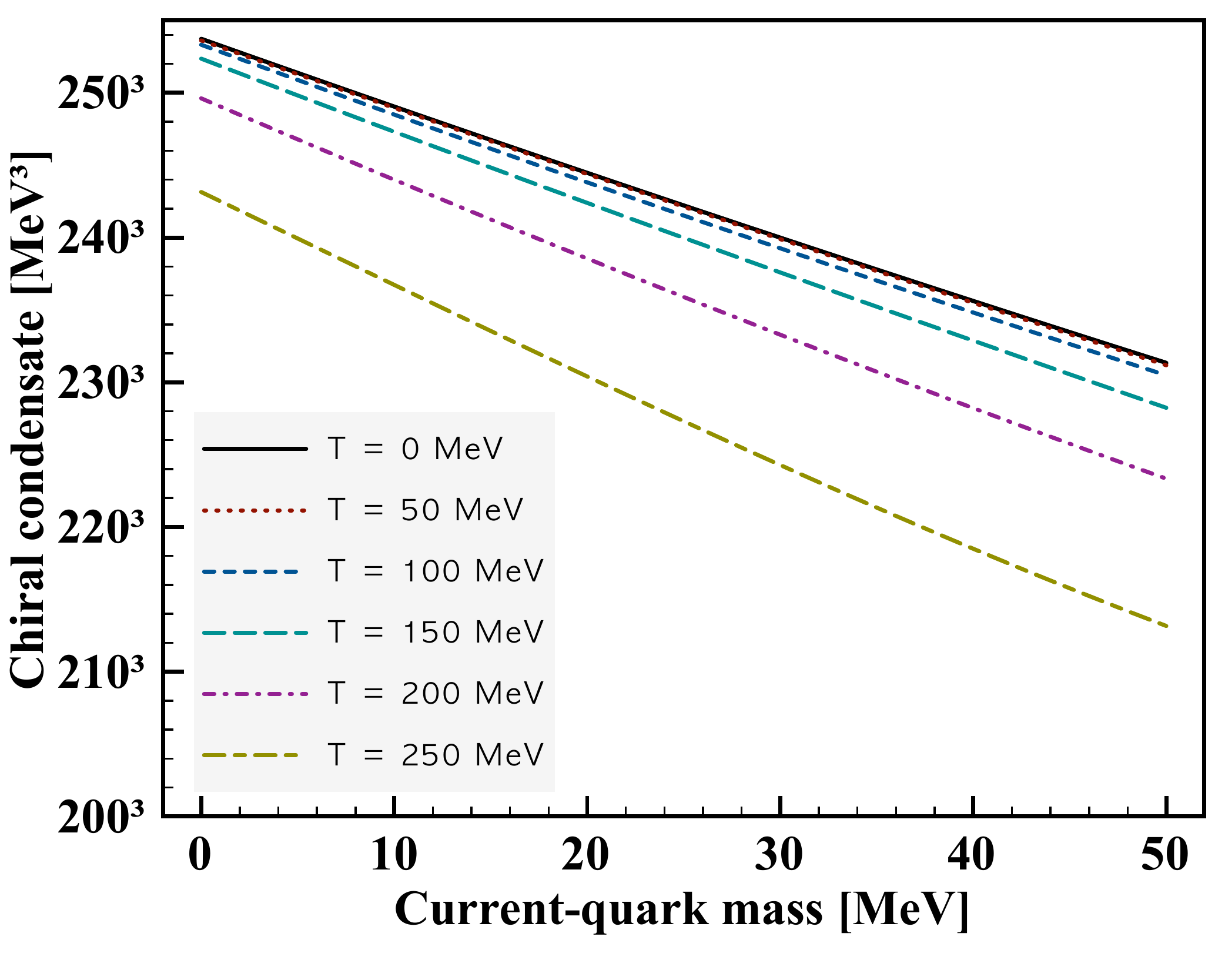}
\includegraphics[width=8.5cm]{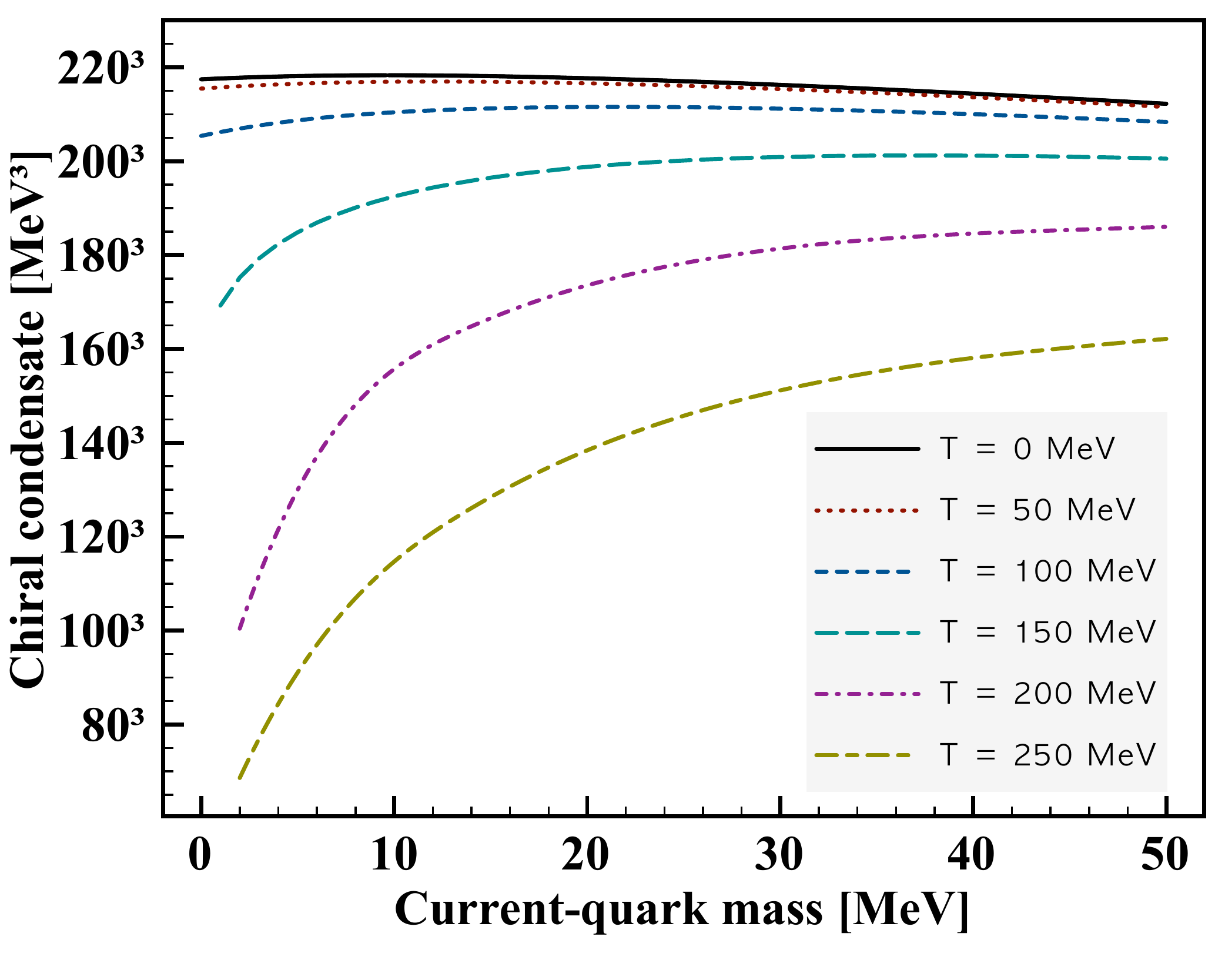}
\end{tabular}
\caption{(Color online) Chiral condensate $\langle iq^{\dagger}q\rangle $ in Eq.~(\ref{eq:CC}) as a function of current-quark mass $m_{q}$ for the different values of temperature, $T=(0\sim250)$ MeV without (left) and with (right) the meson-loop corrections.}
\label{FIG56}
\end{figure}

In Figure~\ref{FIG56}, we show the chiral condensate as a function of $m_{q}$ for $T=(0\sim250)$ MeV without (left) and with (right) MLC. The global behaviors of the curves are very similar to those of $M_{0}$. Without MLC, the chiral condensate decreases linearly with respect to $m_{q}$. Again, we observe the manifestation of the explicit chiral-symmetry breaking with MLC as seen in the right panel of Figure~\ref{FIG56}. In addition, for the case with MLC, we see that the slope of the curves at $m_{q}\approx0$ increases then decreases around $T=200$ MeV. In other words, at a certain value of $T$ between $150$ MeV and $250$ MeV, there appears an inflection point with respect to $T$. Considering that this slope corresponds to chiral susceptibility~\cite{Nam:2008bq}, the inflection point indicates the chiral phase transition at $T^{\chi}_{c}$. We will see this in detail in what follows.

\begin{figure}[t]
\begin{tabular}{cc}
\includegraphics[width=8.5cm]{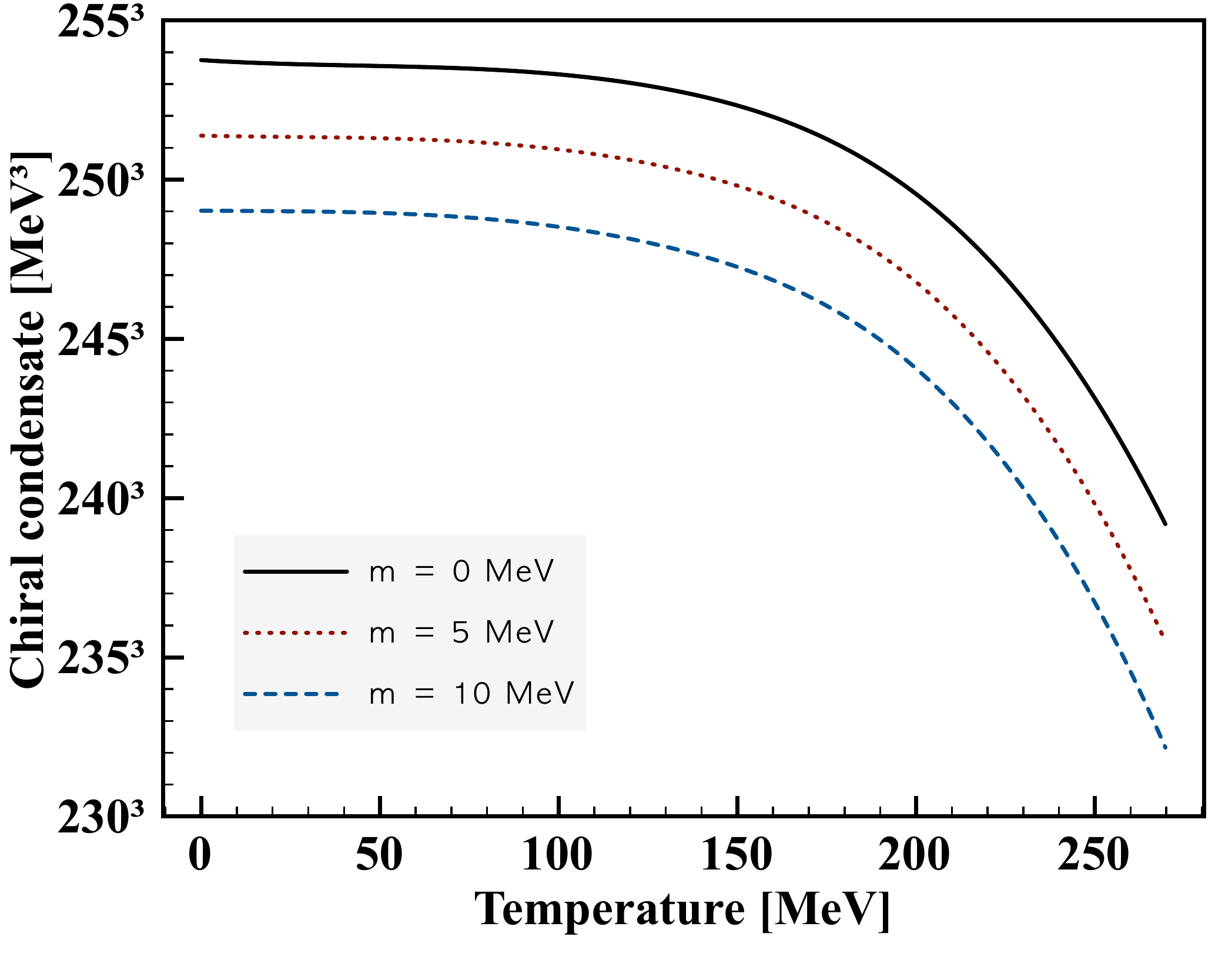}
\includegraphics[width=8.5cm]{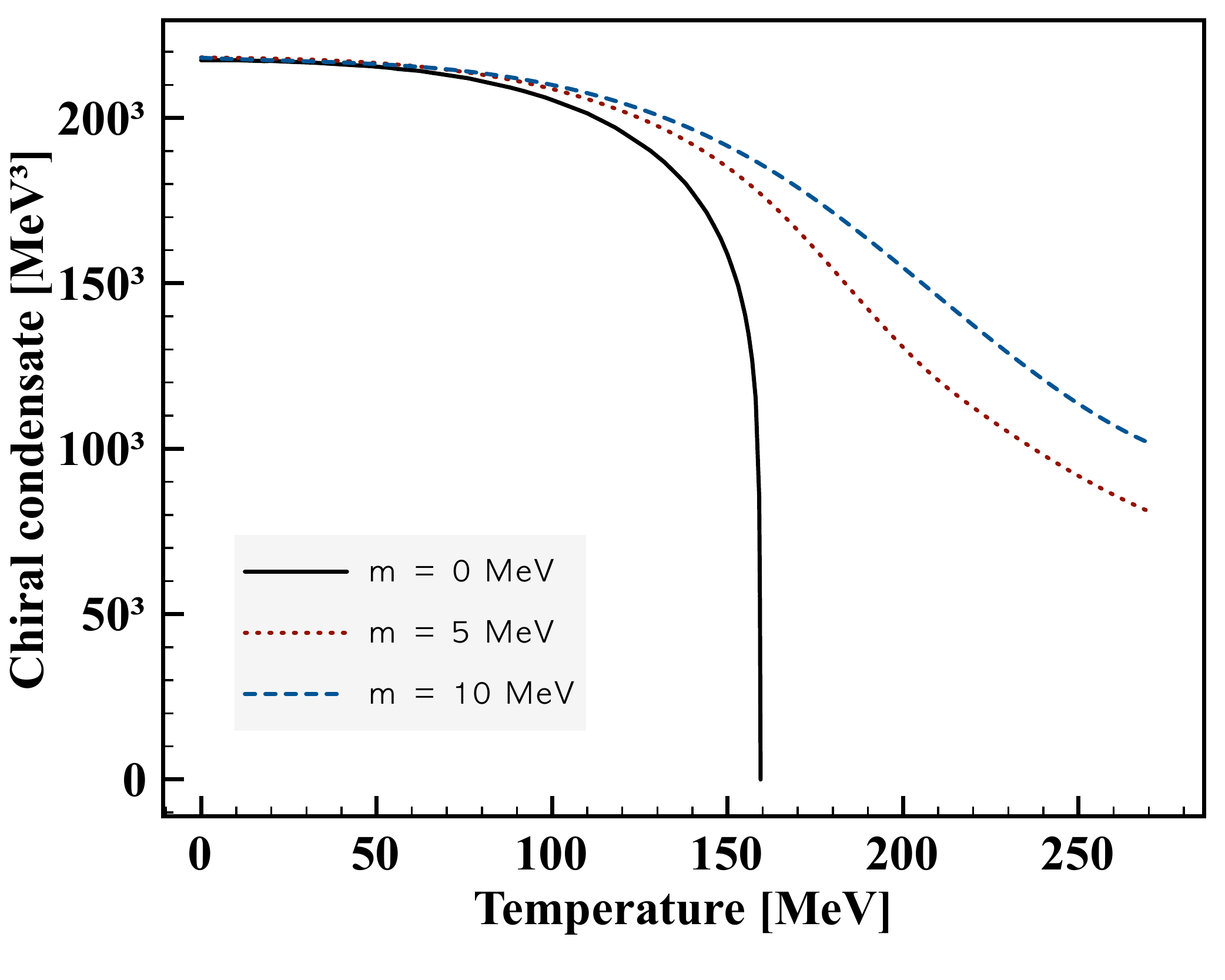}
\end{tabular}
\caption{(Color online) Chiral condensate $\langle iq^{\dagger}q\rangle $ in Eq.~(\ref{eq:CC}) as a function of $T$ for the different values of $m=(0\sim10)$ MeV without (left) and with (right) the meson-loop corrections.}
\label{FIG78}
\end{figure}

The numerical results for the chiral condensate are drawn in Figure~\ref{FIG78} as a function of $T$ for $m_{q}=(0,5,10)$ MeV. Since the empirical values for $u$ and $d$ quarks lie below about $10$ MeV, $m_{u}=(2.55^{+0.75}_{-1.05})$ MeV and $m_{d}=(5.04^{+0.96}_{-1.54})$ MeV for instance~\cite{Amsler:2008zzb}, we confined ourselves to small $m_{q}$ values. In the  left panel of Figure~\ref{FIG78}, the chiral condensate without MLC shows smoothly decreasing curves, and the magnitude  decreases for larger $m_{q}$ as expected from the left panel of Figure~\ref{FIG56}. These smoothly decreasing curves indicate that the chiral restoration without MLC is crossover with different $m_{q}$ values. This observation is far different from the universality consideration of the chiral phase transition~\cite{Muller:2010am,ShTEXT}, which is second order for $m_{q}=0$ and crossover for $m_{q}\ne0$. This flaw is cured by inclusion of MLC as shown in the right panel of the figure. The solid line in the right panel of Figure~\ref{FIG78} indicates the chiral condensate for $m_{q}=0$, i.e. in the chiral limit, whereas the dot and dash lines those for $m_{q}=5$ MeV and $10$ MeV, respectively. Obviously, in the chiral limit, we observe the second-oder chiral phase transition, and it becomes crossover for $m_{q}\ne0$, just satisfying the universality class of the chiral phase transition~\cite{Pisarski:1983ms}. Moreover, the differences between the curves are negligible for the region below $T\lesssim100$ MeV in contrast to those without MLC. This behavior is consistent with that shown in the right panel of Figure~\ref{FIG56}: The chiral condensate is insensitive to $m_{q}$ below $T\approx100$ MeV. We note that these results are consistent in principle with those given in Ref.~\cite{Muller:2010am}, in which NJL model was exploited beyond the mean-field approximation.  However, here we would like to give some comments concerning this universality-class pattern of the chiral restoration via the present framework with the MLC: 1) Although the MLC contribution is certainly very important and plays a dominant role among the $1/N_{c}$ corrections, there are still other $1/N_{c}$ corrections which are not included in our calculation. Consequently one cannot exclude the possibility that the result may be different with the full consideration including the complete $1/N_{c}$ corrections. 2) We modified the instanton parameters using the the Harrington-Shepard caloron, which ignores the dynamical quark contributions. The outcome would be different if the different types of caloron solutions such as the KvBLL one are used. 
(see, Refs~\cite{Diakonov:2005qa,Diakonov:2004jn,Diakonov:2008sg} for recent developments based on the calorons with the quark determinant.). 3) We replace the fourth component of the pion momentum $q$ by the cutoff mass $\Lambda\propto m_{\pi}$ as in Eq.~(\ref{eq:LAMBDA}). Although we verified that the change of the value of the $\Lambda$ does not cause qualitatively significant changes in the relevant physical quantities, the $T$ dependence of the $\Lambda$ ($\Lambda\propto\bar{\rho}_{0}/\bar{\rho}$) would be too simple to be real.
Even it satisfies the $T$ dependence of the $m_{\pi}$: the $m_{\pi}$ increases as the SB$\chi$S is partially restored. 
Naturally a more realistic $T$ dependence of the $\Lambda$ would produce different outcome.
Admittedly our present results may be altered to a certain extend due to these ingredients mentioned above. 
Nevertheless, since either the full $1/N_{c}$ calculation or the similar calculation using the KvBLL caloron is formidable task 
which requires much greater effort and time to achieve. On the other hand, 
our relative simple model calculation did grasp the effect of the MLC contribution which is believed to be dominant one
, we consider that our current result is still helpful and very
instructive for the further development of understanding the phase structure of QCD.
We would like to leave the more sophisticated approach for the future works.

The critical $T$ for the chiral phase transition, $T^{\chi}_{c}$ can be obtained by differentiating the chiral condensate with respect to $T$ as mentioned in the previous Section, resulting in $T^{\chi}_{c}=159$ MeV in the chiral limit, as long as the phenomenological choices of the instanton parameters $\bar{R}\approx1$ fm and $\bar{\rho}\approx1/3$ fm at $T=0$ are adopted. As for $m_{q}=(5,10,15)$ MeV, we obtain $T^{\chi}_{c}=(177,186,196)$ MeV, respectively. According to this observation, $T^{\chi}_{c}$ increases as $m_{q}$ does. It is worth mentioning that, from LQCD simulations in the chiral limit, it turned out that $T^{\chi}_{c}\approx180$ MeV for $N_{f}=2$ using the clover-improved Willson fermions~\cite{Maezawa:2007fd}. Also, using the renormalization-group (RG) improved action, it was found that $T^{\chi}_{c}\approx171$ MeV~\cite{Ali Khan:2000iz}. In Ref.~\cite{de Forcrand:2002ci}, employing the $N_{f}=2$ staggered fermions, the critical $T$ was estimated as $T^{\chi}_{c}=(165\sim181)$ MeV. From effective-model approaches, Ref.~\cite{Blank:2010bz}, using the Schwinger-Dyson model, obtained $T^{\chi}_{c}$ in a wide ranges depending on model parameters: $T^{\chi}_{c}=(82\sim169)$ MeV. Similarly, the NJL model analysis showed $T^{\chi}\approx125$ MeV with the $1/N_{c}$ next-to-leading order (NLO) computation~\cite{Muller:2010am}. In their work, it was also shown that $T^{\chi}_{c}$ appears at higher $T$ without the NLO contribution, i.e. in the mean-field approximation. However, even in the case without the NLO contribution, the pattern of chiral symmetry restoration presents second-order phase transition, being different from ours.

Comparing with the LQCD values, our current value ($T^{\chi}_{c}=159$ MeV) presents only about $5\sim20\%$ deviation. Hence, we can conclude that, qualitatively, the assumptions discussed in Sections IV and V can be justified. We also note that $T^{\chi}_{c}$ can be much closer to the LQCD values by tuning the instanton parameters within possible changes of $\bar{R}$ and $\bar{\rho}$ as in Ref.~\cite{Goeke:2007bj,Chu:1994vi,Negele:1998ev}. We obtain $T^{\chi}_{c}=(168\sim177)$ MeV by choosing $\bar{R}=(0.98\sim0.96)$ fm. It is also observed that $T^{\chi}_{c}$ is very sensitive to $\bar{R}$. In other words, it is sensitive to the diluteness of the instanton ensemble. Technically, this tendency can be understood by that $\bar{R}$ controls the instanton-number density $N/V\approx1/\bar{R}^{4}$, which determines all the strength of relevant quantities in the present framework as in Eq.~(\ref{eq:NOV}). Theoretically, it is interesting to see that $T^{\chi}_{c}$ is inversely proportional to $\bar{R}$ as shown above. We can explain this as follows: If the instanton ensemble is relatively denser, corresponding to smaller $\bar{R}$, the vacuum effect tends to remain effective up to higher $T$. In contrast, in the dilute instanton ensemble, the chiral NG phase which is governed by the vacuum effect will be terminated more easily at lower $T$. In this way, we can understand the higher (lower) $T^{\chi}_{c}$ for the dense (dilute) instanton ensemble, respectively. Numerical results are summarized in Table~\ref{TABLE2}.

\begin{figure}[t]
\includegraphics[width=10cm]{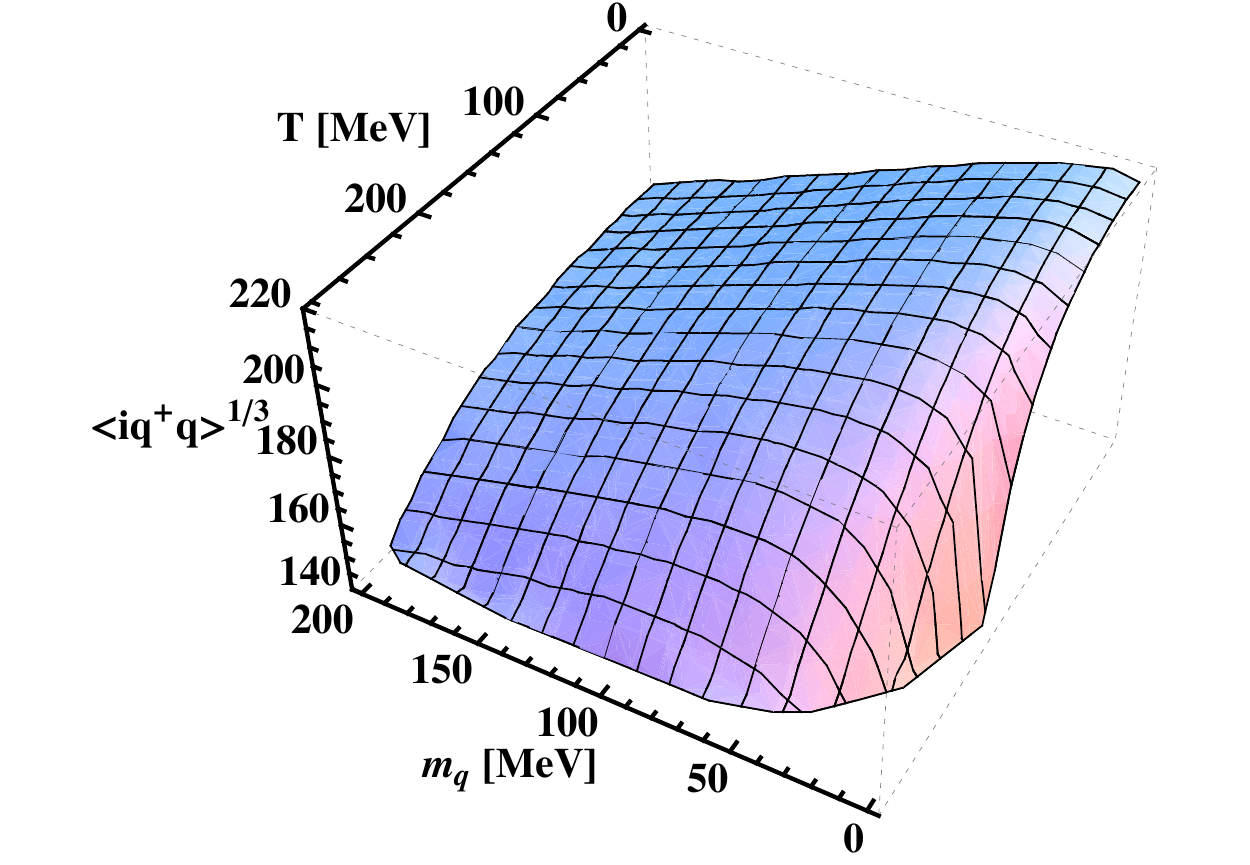}
\caption{(Color online)  Chiral condensate $\langle iq^{\dagger}q\rangle^{1/3}$ [MeV] in Eq.~(\ref{eq:CC}) as a function of $T$ and $m_{q}$.}
\label{FIG9}
\end{figure}

Finally, in Figure~\ref{FIG9}, we plot the chiral condensate as a function of $T$ and $m_{q}$. Although we have been interested only in the SU(2) light-flavor sector, we  show the numerical result up to $m_{q}\approx200$ MeV to see overall behaviors for the $m_{q}$ dependence of the chiral condensate. It is clear that the second-order chiral restoration at $m_{q}=0$ turns into that of crossover gradually as $m_{q}$ increases. In addition, the magnitude of the chiral condensate decreases as $m_{q}$ becomes more massive. This is a consequence of general behaviors of the QCD vacuum contribution which decreases as $m_{q}$ increases~\cite{Nam:2006ng}. The decreasing instanton effects make the chiral condensate smaller as $T$ gets higher as already shown in Figure.~\ref{FIG78}.

\begin{table}[b]
\begin{tabular}{c||c|c|c|c}
$m_{q}$&$0$ MeV&$5$ MeV&$10$ MeV &$15$ MeV\\
\hline
Phase transition&Second order& Crossover&Crossover&Crossover\\
\hline
$T^{\chi}_{c}$&$159$ MeV&$177$ MeV&$186$ MeV&$196$ MeV
\end{tabular}
\caption{$T^{\chi}_{c}$ and phase-transition pattern for different values of current-quark masses.}
\label{TABLE2}
\end{table}

\section{Summary and conclusion}
We have studied the pattern of chiral restoration for the SU(2) light-flavor sector at finite $T$ beyond the chiral limit.
We used the instanton-vacuum configuration in the grand-canonical instanton ensemble. The relevant instanton parameters such as $\bar{R}$ and $\bar{\rho}$ were modified at finite $T$ by using the Harrington-Shepard caloron. As have been noticed in the previous works, MLC, which correspond to the $1/N_{c}$ corrections, are critical ingredients to incorporate appropriate $m_{q}$ dependences of the physical observables. Hence, we employed MLC in the present work to investigate the pattern of chiral symmetry restoration
as a function of $m_{q}$. We computed the effective quark mass and the chiral condensate as functions of $T$ and/or $m_{q}$. The critical $T$ for the chiral-phase transition, $T^{\chi}_{c}$ was obtained by differentiating the chiral condensate with respect to $T$.
Our main results are listed below:
\begin{itemize}
\item As a general consequence in the present framework, the instanton contribution is weakened as $T$ increases because the instanton ensemble becomes dilute. It results in the decreasing magnitudes of the chiral order parameters. As a result, the magnitude of the effective quark mass as well as the chiral condensate, as the order parameters for the chiral-phase transition, get diminished with respect to $T$. Inclusion of MLC does not change this trend.
\item In contrast, their $m_{q}$ dependences are largely dependent on whether MLC is taken into account or not. Without MLC, the chiral order parameters are monotonically decreasing functions with respect to $m_{q}$. If and only if MLC is taken into account, the computed effective quark mass can be comparable with that from LQCD simulations.
\item MLC is indeed responsible for appropriate chiral-restoration patterns, which follow the universality-class restoration patterns depending on $m_{q}$: Second-order and crossover phase transitions for $m_{q}=0$ and $m_{q}>0$, respectively, for $N_{f}=2$. Without MLC the chiral-restoration patterns are all crossover, regardless of the value of $m_{q}$.
\item Our results show that $T^{\chi}_{c}=(159,177,186,196)$ MeV for $m_{q}=(0,5,10,15)$ MeV for the SU(2) light-flavor sector, when we employ the phenomenological choices for the instanton parameters, $\bar{R}\approx1$ fm and $\bar{\rho}\approx1/3$ fm at $T$=0. We observe that $T^{\chi}_{c}$ is sensitive to $\bar{R}$, and the LQCD compatible values of $T^{\chi}_{c}$ can be obtained by choosing the instanton parameters, which are deviated by a few percent from the phenomenological ones.
\end{itemize}
Overall, we conclude that MLC is crucial to produce correct patterns of chiral restoration for $N_{f}=2$ in the present framework. This observation is very important when one tries to apply this framework to study the QCD phase diagram, in particular, the positions of CEP and TCP. The extension of the present results to the nonzero $\mu$ and/or $N_{f}=2+1$ cases are urgent tasks to explore realistic understanding of QCD in medium. The related works are underway.

\section*{Acknowledgment}
We thank C.~-J.~David~Lin and B.~G.~Yu for useful discussions.
S.i.N was supported by the grant NSC 98-2811-M-033-008 and
C.W.K  was supported by the grant NSC 96-2112-M-033-003-MY3 from National Science Council (NSC) of Taiwan. The support from National Center for Theoretical Sciences (North) of Taiwan (under the grant number NSC 97-2119-M-002-001) is also acknowledged. The work of S.i.N. was also partially supported by the grant NRF-2010-0013279 from National Research Foundation (NRF) of Korea.
\section*{Appendix}
The relevant functions in Eqs.~(\ref{eq:NOV}) and (\ref{eq:CC}) are defined as follows:
\begin{eqnarray}
\label{eq:RFUNC}
F_{0}&=&T\sum_{n=-\infty}^{\infty}
\frac{M_{a}\bar{M}_{a}}{w^{2}_{n}+E^{2}_{a}}
=\frac{M_{a}\bar{M}_{a}}{2E_{a}}\mathrm{tanh}\left(\frac{E_a}{2T}\right),
\cr
F_{1}&=&T\sum_{n=-\infty}^{\infty}
\frac{3M_{a}M_{b}w^{2}_{n}}{(w^{2}_{n}+E^{2}_{a})(w_{n}^{2}+E^{2}_{b})}
=\frac{3M_{a}M_{b}}{2(E^{2}_{a}-E^{2}_{b})}
\left[E_{a}\mathrm{tanh}\left(\frac{E_a}{2T}\right)
-E_{b}\mathrm{tanh}\left(\frac{E_b}{2T}\right) \right],
\cr
F_{2}&=&T\sum_{n=-\infty}^{\infty}
\frac{3M_{a}M_{b}\xi^{2}}{(w^{2}_{n}+E^{2}_{a})(w_{n}^{2}+E^{2}_{b})}
=\frac{3M_{a}M_{b}\xi^{2}}{2E_{a}E_{b}(E^{2}_{a}-E^{2}_{b})}
\left[E_{a}\mathrm{tanh}\left(\frac{E_b}{2T}\right)-E_{b}\mathrm{tanh}
\left(\frac{E_a}{2T}\right) \right],
\cr
G_{0}&=&\frac{\bar{M}_{a}}{2E_{a}}\mathrm{tanh}\left(\frac{E_{a}}{2T}\right),
\,\,\,\,
G_{1}=\frac{m}{2E_{0}}\mathrm{tanh}\left(\frac{E_{a0}}{2T}\right),
\cr
G_{2}&=&T\sum_{n=-\infty}^{\infty}
\frac{3M_{a}M_{b}(\bar{M}_{a}+\bar{M}_{b})}
{(w^{2}_{n}+E^{2}_{a})(w_{n}^{2}+E^{2}_{b})}
=\frac{3M_{a}M_{b}(\bar{M}_{a}+\bar{M}_{b})}{2E_{a}E_{b}(E^{2}_{a}-E^{2}_{b})}
\left[E_{a}\mathrm{tanh}\left(\frac{E_b}{2T}\right)-E_{b}\mathrm{tanh}\left(\frac{E_a}{2T}\right) \right],
\end{eqnarray}
where the definitions of $M_{a,b}$, $\bar{M}_{a,b}$ and $E_{a,b}$ are given in the  text.



\begin{thebibliography}{99}
\bibitem{Klingl:1997kf}
  F.~Klingl, N.~Kaiser and W.~Weise,
  Nucl.\ Phys.\  A {\bf 624}, 527 (1997).
\bibitem{Hatsuda:1991ez}
  T.~Hatsuda and S.~H.~Lee,
  Phys.\ Rev.\  C {\bf 46}, 34 (1992).
\bibitem{Kwon:2008vq}
  Y.~Kwon, M.~Procura and W.~Weise,
  Phys.\ Rev.\  C {\bf 78}, 055203 (2008).
\bibitem{Buballa:2003qv}
  M.~Buballa,
  Phys.\ Rept.\  {\bf 407}, 205 (2005).
\bibitem{Muller:2010am}
  D.~Muller, M.~Buballa and J.~Wambach,
  Phys.\ Rev.\  D {\bf 81}, 094022 (2010).
\bibitem{Schwarz:1999dj}
  T.~M.~Schwarz, S.~P.~Klevansky and G.~Papp,
  Phys.\ Rev.\  C {\bf 60}, 055205 (1999).
\bibitem{Blank:2010bz}
  M.~Blank and A.~Krassnigg,
  Phys.\ Rev.\  D {\bf 82}, 034006 (2010).
\bibitem{Hong:1999fh}
  D.~K.~Hong {\it et al.},
  Phys.\ Rev.\  D {\bf 61}, 056001 (2000)
  [Erratum-ibid.\  D {\bf 62}, 059903 (2000)].
\bibitem{Fukushima:2003fw}
  K.~Fukushima,
  Phys.\ Lett.\  B {\bf 591}, 277 (2004).
\bibitem{Ratti:2004ra}
  C.~Ratti and W.~Weise,
  Phys.\ Rev.\  D {\bf 70}, 054013 (2004).
\bibitem{Ratti:2005jh}
  C.~Ratti, M.~A.~Thaler and W.~Weise,
  Phys.\ Rev.\  D {\bf 73}, 014019 (2006).
\bibitem{Ghosh:2006qh}
  S.~K.~Ghosh, T.~K.~Mukherjee, M.~G.~Mustafa and R.~Ray,
  Phys.\ Rev.\  D {\bf 73}, 114007 (2006).
\bibitem{Aharony:2006da}
  O.~Aharony, J.~Sonnenschein and S.~Yankielowicz,
  Annals Phys.\  {\bf 322}, 1420 (2007).
\bibitem{Herzog:2006ra}
  C.~P.~Herzog,
  Phys.\ Rev.\ Lett.\  {\bf 98}, 091601 (2007).
\bibitem{Kobayashi:2006sb}
  S.~Kobayashi, D.~Mateos, S.~Matsuura, R.~C.~Myers and R.~M.~Thomson,
  JHEP {\bf 0702}, 016 (2007).
\bibitem{Harada:2003wa}
  M.~Harada and C.~Sasaki,
  Nucl.\ Phys.\  A {\bf 736}, 300 (2004).
\bibitem{Brown:2009az}
  G.~E.~Brown, M.~Harada, J.~W.~Holt, M.~Rho and C.~Sasaki,
  Prog.\ Theor.\ Phys.\  {\bf 121}, 1209 (2009).
\bibitem{Harada:2003jx}
  M.~Harada and K.~Yamawaki,
  Phys.\ Rept.\  {\bf 381}, 1 (2003).
\bibitem{Kirchbach:1997rk}
  M.~Kirchbach and A.~Wirzba,
  Nucl.\ Phys.\  A {\bf 616}, 648 (1997).
\bibitem{Meissner:2001gz}
  U.~G.~Meissner, J.~A.~Oller and A.~Wirzba,
  Annals Phys.\  {\bf 297} (2002) 27.
\bibitem{GomezNicola:2004gg}
  A.~Gomez Nicola, F.~J.~Llanes-Estrada and J.~R.~Pelaez,
  Phys.\ Lett.\  B {\bf 606}, 351 (2005).
\bibitem{Diakonov:1988my}
  D.~Diakonov and A.~D.~Mirlin,
  Phys.\ Lett.\  B {\bf 203}, 299 (1988).
\bibitem{Carter:1998ji}
  G.~W.~Carter and D.~Diakonov,
  Phys.\ Rev.\  D {\bf 60}, 016004 (1999).
\bibitem{Nam:2008bq}
  S.~i.~Nam,
  Phys.\ Rev.\  D {\bf 79}, 014008 (2009).
\bibitem{Nam:2009nn}
  S.~i.~Nam,
  J.\ Phys.\ G {\bf 37}, 075002 (2010).
\bibitem{Braun:2008pi}
  J.~Braun,
  Eur.\ Phys.\ J.\  C {\bf 64}, 459 (2009).
\bibitem{Braun:2009si}
  J.~Braun,
  Phys.\ Rev.\  D {\bf 81}, 016008 (2010).
\bibitem{Stephanov:1998dy}
  M.~A.~Stephanov, K.~Rajagopal and E.~V.~Shuryak,
  Phys.\ Rev.\ Lett.\  {\bf 81}, 4816 (1998).
\bibitem{Hatta:2002sj}
  Y.~Hatta and T.~Ikeda,
  Phys.\ Rev.\  D {\bf 67}, 014028 (2003).
\bibitem{deForcrand:2006pv}
  P.~de Forcrand and O.~Philipsen,
  JHEP {\bf 0701}, 077 (2007).
\bibitem{Fodor:2004nz}
  Z.~Fodor and S.~D.~Katz,
  JHEP {\bf 0404}, 050 (2004).
\bibitem{de Forcrand:2002ci}
  P.~de Forcrand and O.~Philipsen,
  Nucl.\ Phys.\  B {\bf 642}, 290 (2002).
\bibitem{Schaefer:2004en}
  B.~J.~Schaefer and J.~Wambach,
  Nucl.\ Phys.\  A {\bf 757}, 479 (2005).
\bibitem{Schaefer:2006ds}
  B.~J.~Schaefer and J.~Wambach,
  Phys.\ Rev.\  D {\bf 75}, 085015 (2007).
\bibitem{Schafer:1996wv}
  T.~Schafer and E.~V.~Shuryak,
  Rev.\ Mod.\ Phys.\  {\bf 70}, 323 (1998).
\bibitem{Diakonov:2002fq}
  D.~Diakonov,
  Prog.\ Part.\ Nucl.\ Phys.\  {\bf 51}, 173 (2003).
\bibitem{Goeke:2007bj}
  K.~Goeke, M.~M.~Musakhanov and M.~Siddikov,
  Phys.\ Rev.\  D {\bf 76}, 076007 (2007).
\bibitem{Kim:2005jc}
  H.~C.~Kim, M.~M.~Musakhanov and M.~Siddikov,
  Phys.\ Lett.\  B {\bf 633}, 701 (2006).
\bibitem{Bowman:2005vx}
  P.~O.~Bowman {\it et al.},
  Phys.\ Rev.\  D {\bf 71}, 054507 (2005).
\bibitem{Harrington:1976dj}
  B.~J.~Harrington and H.~K.~Shepard,
  Nucl.\ Phys.\  B {\bf 124}, 409 (1977).
\bibitem{Kraan:1998pm}
  T.~C.~Kraan and P.~van Baal,
  Nucl.\ Phys.\  B {\bf 533}, 627 (1998).
\bibitem{Lee:1998bb}
  K.~M.~Lee and C.~h.~Lu,
  Phys.\ Rev.\  D {\bf 58}, 025011 (1998).
\bibitem{Nowak:1989jd}
  M.~A.~Nowak, J.~J.~M.~Verbaarschot and I.~Zahed,
  Nucl.\ Phys.\  B {\bf 325}, 581 (1989).
\bibitem{Rossner:2007ik}
  S.~Roessner, T.~Hell, C.~Ratti and W.~Weise,
  Nucl.\ Phys.\  A {\bf 814}, 118 (2008).
\bibitem{Musakhanov:2001pc}
  M.~Musakhanov,
  arXiv:hep-ph/0104163.
\bibitem{Novikov:1981xi}
  V.~A.~Novikov, M.~A.~Shifman, A.~I.~Vainshtein and V.~I.~Zakharov,
  Nucl.\ Phys.\  B {\bf 191}, 301 (1981).
\bibitem{Amsler:2008zzb}
  C.~Amsler {\it et al.}  [Particle Data Group],
  Phys.\ Lett.\  B {\bf 667} 1 (2008).
\bibitem{ShTEXT}
E.~V.~Shuryak, The QCD vacuum, hadrons and superdense matter (World Scientific, Singapore, 1988).
\bibitem{Pisarski:1983ms}
  R.~D.~Pisarski and F.~Wilczek,
  Phys.\ Rev.\  D {\bf 29}, 338 (1984).
\bibitem{Diakonov:2005qa}
  D.~Diakonov and N.~Gromov,
  Phys.\ Rev.\  D {\bf 72}, 025003 (2005).
\bibitem{Diakonov:2004jn}
  D.~Diakonov, N.~Gromov, V.~Petrov and S.~Slizovskiy,
  Phys.\ Rev.\  D {\bf 70}, 036003 (2004).
\bibitem{Diakonov:2008sg}
  D.~Diakonov,
  Acta Phys.\ Polon.\  B {\bf 39}, 3365 (2008).
\bibitem{Maezawa:2007fd}
  Y.~Maezawa {\it et al.},
  J.\ Phys.\ G {\bf 34}, S651 (2007).
\bibitem{Ali Khan:2000iz}
  A.~Ali Khan {\it et al.}  [CP-PACS Collaboration],
  Phys.\ Rev.\  D {\bf 63}, 034502 (2001).
\bibitem{Chu:1994vi}
  M.~C.~Chu, J.~M.~Grandy, S.~Huang and J.~W.~Negele,
  Phys.\ Rev.\  D {\bf 49}, 6039 (1994).
\bibitem{Negele:1998ev}
  J.~W.~Negele,
  Nucl.\ Phys.\ Proc.\ Suppl.\  {\bf 73}, 92 (1999).
\bibitem{Nam:2006ng}
  S.~i.~Nam and H.~C.~Kim,
  Phys.\ Lett.\  B {\bf 647}, 145 (2007).
\end{thebibliography}
\end{document}